\documentclass[twocolumn,showpacs,preprintnumbers,amsmath,amssymb]{revtex4}

\usepackage[latin1]{inputenc}
\usepackage{makeidx}
\usepackage{layout}        
\usepackage{color}
\usepackage{graphicx}
\usepackage{dcolumn}
\usepackage{bm}
\usepackage{float}

\newcommand{\bra}[1]{\langle #1|}
\newcommand{\ket}[1]{|#1\rangle}

\begin{document}
\preprint{preprint}

\title{Electronic transport in iron atomic contacts: from the infinite wire to realistic geometries. }

\author{   Gabriel Aut\`es$^*$, Cyrille Barreteau$^*$, Daniel Spanjaard$^{\dag}$
and Marie-Catherine Desjonqu\`eres$^*$ }
\affiliation{$^*$CEA Saclay, IRAMIS/SPCSI, B\^atiment 462, F-91191 Gif sur
Yvette, France }

\affiliation{$^{\dag}$Laboratoire de Physique des Solides,
             Universit\'e Paris Sud, B\^atiment 510, F-91405 Orsay, France}

\date{\today}
\begin{abstract}
We present a theoretical study of spin polarized transport in Fe
atomic contacts using a self-consistent tight-binding Hamiltonian in a
non-orthogonal $s$, $p$ and $d$ basis set, the spin-polarization being obtained
from a non-collinear Stoner-like model and the transmission probability from the Fisher-Lee formula.
The behaviour of an infinite perfect Fe wire is
compared with that of an infinite chain presenting geometric defects or magnetic walls
and with that of a finite chain connected to infinite one-dimensional or three-dimensional
leads. In the presence of defects or contacts the transmission probability of $d$
electrons is much more affected than that of $s$ electrons, in particular, contact effects may
suppress some transmission channels. It is shown that the behaviour of an infinite
wire is never obtained even in the limit of long chains connected to electrodes.
The introduction of the spin-orbit coupling term in the Hamiltonian enables us
to calculate the anisotropy of the magneto-resistance.
Finally whereas the variation of the magneto-resistance as a function of
the magnetization direction is step-like for an infinite wire, 
it becomes smooth in the presence of defects or contacts. 
\end{abstract}

\maketitle

\section{Introduction}

The electronic transport properties of atomic point contacts between two metallic electrodes
have recently been the subject of great interest, both from the experimental and
theoretical points of view. On the experimental side such atomic sized conductors
can be obtained either by means of scanning tunneling microscopes \cite{Ohnishi98} or by using
mechanically controllable break junctions \cite{Muller92}. In some materials, for instance gold, monatomic
chains, several atom long, are spontaneously obtained \cite{Ohnishi98,Yanson98}.

In these experiments the conductance is measured during the elongation process and shows clear 
plateaus corresponding to the various stable atomic arrangements in the constriction
region. In noble and alkali metals the last plateaus before breaking
are  very close to multiples of $2e^2/h$ \cite{Yanson98}. Oscillations of the conductance
with the length of the suspended chain are also observed \cite{Smit03}. In transition metals 
even though plateaus are still  present, the quantized behaviour
is much less clear. However, thanks to the great mechanical
stability of the break junction technique,  the system can be stabilized
for a given value of the conductance. This is essential 
in the case of magnetic materials since it allows to perform a series
of magneto-resistive measurements revealing a remarkably large
anisotropic magnetoresistance (AMR) effects in materials like iron, cobalt or nickel
\cite{Viret02,Viret06}.

Understanding these transport properties is a challenging problem and is of prime importance
for their future applications in nano and spin electronics. Theoretically this relies
on the calculation of the transmission factor $T(E)$ of an electron at energy $E$
from an electrode to the other in the ballistic regime which is relevant for such devices.
 In a perfectly periodic nanowire the transmission factor of each electronic
state is unity. Therefore at small bias voltage the conductance is determined
by the number of linearly independent electronic states at the Fermi level propagating
in a given direction (transmission channels), and is  quantized in units
of $2e^2/h$ or $e^2/h$ for a non-magnetic or magnetic material, respectively \cite{Datta}.

In a realistic system the electrodes are macroscopic and connected
by an atomic sized constriction region. For such a system $T(E)$ is limited
by the number of channels in the narrowest part of the constriction. However, the
transmission factor is usually smaller than this number due to scattering effects at the
edges of the system. Several methods have been proposed to calculate the electronic
transport in atomic contacts but the most popular one is using an
expansion of the electronic states on a local basis set
which allows an easy partition of the system into three parts: the two leads and a central part.
In this approach $T(E)$ can conveniently be obtained
from the Green function of the central part interacting with the electrodes, by
using the Fisher-Lee formula \cite{Fisher81,Viljas05,Sanvito06,Thygesen06}.

In this work we use a tight-binding Green function formalism to investigate the role of
(geometric or magnetic) defects and contacts on the transmission factor, and thus
the conductance at low-bias voltages, of atomic junctions with simple geometries.
Most of the calculations we present here have been devoted to iron, using a realistic tight-binding
Hamiltonian \cite{Mehl96} which includes magnetism and, possibly, 
the spin-orbit coupling term \cite{Autes06}. For the sake
of comparison we have also considered gold which contrary to iron is non-magnetic
and has electronic states of pure $s$ character at the Fermi level.

We show in the following that the monatomic perfect wire model \cite{Ono03,Velev05,Viret06,Sokolov07} 
which has been used to explain experimental results is not reliable, at least for transition metals,
since the presence of defects and contacts strongly perturbs the transmission
factor of the various conductance channels. Indeed, even in the limit of long chains,
the behaviour of the perfect infinite wire is not recovered due to contact effects.

In Sec. II we present the theoretical model and give some details about practical computations.
Sec. III A is devoted to the study of the influence of a geometrical or magnetic defect in a
monatomic iron wire. The role played by atomic contacts is discussed in Sec. III B, and a comparison
is made between Fe and Au. Conclusions are drawn in Sec. IV. Finally details of algebraic calculations
are presented in the appendices. In appendix A we derive the Fisher-Lee formula for the general
case in which the atomic spin-orbitals (and thus the overlap integrals) and the Hamiltonian matrix
may be complex. Appendix B explains the method used for the calculation of the surface Green function.
Appendix C gives the main steps in the derivation of the transmission factor for a simple model studied in the  text.

\section{Method}

\subsection{The tight-binding model}

The electronic structure is derived from an Hamiltonian operator $\hat{H}$ 
expressed in a non-orthogonal basis set of $s$, $p$ and $d$  atomic spin-orbitals 
hereafter denoted as $\ket{i\lambda\sigma}$ where $i$ is an atomic site,
$\lambda$ an atomic orbital and $\sigma$ the spin.
The matrix elements $\bra{i \lambda \sigma }\hat{H}\ket{j \mu \sigma'}$ of the operator $\hat{H}$
in the $\ket{i \lambda \sigma}$ basis  form the matrix $\bm{H}$.
This matrix is written as the sum of four terms:

\begin{equation}
\bm{H} = \bm{H}_{\text{TB}} + \bm{H}_{\text{ee}} + \bm{H}_{\text{LCN}} + \bm{H}_{\text{so}},
\label{TB}
\end{equation}

$\bm{H}_{\text{TB}}$ is a tight-binding Hamiltonian expressed in terms of Slater-Koster parameters
which are parametrized by a fit on ab-initio calculations for the bulk in the
 non-magnetic state \cite{Mehl96}. 
$\bm{H}_{\text{ee}}$ accounts for the spin-polarization and
is written as a simple intra-atomic Stoner exchange potential applied on $d$
electrons $- I/2 \sum_i \bm{M}_{id}.\bm{\sigma}$, where $\bm{M}_{id}$ is the
$d$ net spin magnetic moment vector of site $i$ and $\bm{\sigma}$ the Pauli matrix vector.
Note that this exchange potential allows for a non collinearity of spins.
 $I$ is the Stoner parameter which we take equal to 1eV as in our previous studies on
 iron \cite{Autes06,Desjonqueres07}.
$\bm{H}_{\text{LCN}}$ is added  to ensure quasi local charge neutrality 
in systems with inequivalent atoms \cite{Autes06}. Its matrix elements 
$\bra{i \lambda \sigma}H_{\text{LCN}}\ket{j \mu \sigma'}$ are written 
$\lambda^{\text{pen}}(\Delta q_i+\Delta q_j)S_{ij}^{\lambda \mu} \delta_{\sigma \sigma'}$, 
$\lambda^{\text{pen}}$ is the penalization factor (in practice $\lambda^{\text{pen}}=2.5$eV),
$\Delta q_i$ is the deviation of the Mulliken charge of site $i$ from the valence
charge, and $S_{ij}^{\lambda \mu}\delta_{\sigma \sigma'}$ are the matrix elements of the overlap matrix
$\bm{S}$. Finally $\bm{H}_{\text{so}}$ is the spin-orbit coupling (SOC) Hamiltonian written as 
$\xi \bm{L}.\bm{\mathcal S}$ where $\bm{L}$  and $\bm{\mathcal S}=\hbar\bm{\sigma}/2$
are the orbital and spin momentum operators, respectively.
Only intra-atomic matrix elements  between $d$ spin orbitals are retained and $\xi$ is 
the spin-orbit coupling constant.
It has been determined for Fe in ref \cite{Autes06} and is equal to $0.06$eV 
when  $\bm{L}$ and $\bm{\mathcal S}$ are expressed in Bohr magnetons.

The electronic (and magnetic) structure of a given system and its corresponding Hamiltonian
are obtained by a self-consistent procedure since $\bm{H}_{\text{ee}}$ and $\bm{H}_{\text{LCN}}$
depend on the charges. 

This model has been checked on bulk, surfaces and monatomic wire of magnetic iron,
and good agreement with \textit{ab initio} calculations \cite{Autes06} was found. 
Finally let us mention that, although we have shown in recent publications
that orbital polarization effects \cite{Desjonqueres07,Desjonqueres07epj} 
(taken into account by a more accurate expression
of the electronic interaction Hamiltonian $\hat{H}_{\text{ee}}$) may have an influence on the electronic
and magnetic properties of nanostructures, we have ignored them in the following
since the aim of the present work is to discuss the effect of contacts on transport properties
that are present whatever $\bm{H}_{\text{ee}}$.

In the following, for the purpose of deriving qualitative arguments, we have also
used an Hamiltonian $\bm{H}$ limited to $\bm{H}_{\text{TB}}$ for an $s$ band.

\subsection{Electronic transport formalism \label{sec:transport}}

We calculate the electric conductance in the Landauer-B\"uttiker approach \cite{Datta}. 

\begin{figure}[!fht]
\begin{center}
\includegraphics*[width=0.8\linewidth,angle=0]{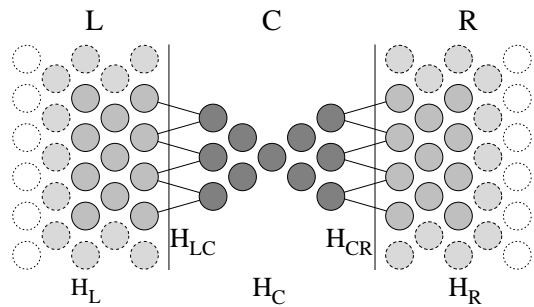}
\end{center}
\caption{Schematic atomic model for the transmission calculation in the Landauer
approach. A scattering region (C) is coupled to two semi-infinite leads (L and
R), L and R are assumed to interact only through C. }
\label{fig:system}
\end{figure}

The system is divided into three parts (Fig. \ref{fig:system}): a central
scattering region (C) and two semi-infinite leads (L and R) with
two-dimensional periodicity.

The transmission probability $T(E)$ of an electron with energy $E$
is obtained from the Green function formalism. Our
approach is close to the one proposed by Sanvito {\sl et al.} \cite{Sanvito06}.
The scattering region in Fig. \ref{fig:system} is chosen wide enough so that no
direct interactions remain between the two leads. The complete Hamiltonian and overlap
 matrices of our system are then

\begin{equation}
\bm{H} = \left( \begin{array}{ccc}
\bm{H}_{L}         & \bm{H}_{LC}        & 0             \\
\bm{H}_{LC}^{\dag} & \bm{H}_{C}         & \bm{H}_{CR} \\
0                  & \bm{H}_{CR}^{\dag} & \bm{H}_{R}         \\
\end{array}\right) 
\quad \bm{S} = \left( \begin{array}{ccc}
\bm{S}_{L}         & \bm{S}_{LC}        & 0             \\
\bm{S}_{LC}^{\dag} & \bm{S}_{C}         & \bm{S}_{CR} \\
0                  & \bm{S}_{CR}^{\dag} & \bm{S}_{R}         \\
\end{array}\right) \\
\label{hamiltonian}
\end{equation}

The matrix $\bm{H}_{C}$ is of size $N_{C}\times N_{C}$ where $N_{C}$ is the number of atomic
spin-orbitals in the central region.
$\bm{H}_{L}$ and $\bm{H}_{R}$ are semi-infinite. The number of non-zero elements of the
coupling and overlap matrices $\bm{H}_{\alpha C}$ and $\bm{S}_{\alpha C}$ is $N_{\alpha} \times N_{C}$ 
where $N_{\alpha}$ is the number of spin-orbitals on atoms in lead $\alpha$ ($L$ or $R$) 
that have hopping and overlap integrals with the central region.

Here we have considered
the general case where the overlap matrix (thus the atomic spin-orbitals), and the Hamiltonian
matrix elements may be complex.
The expression of $T(E)$ is derived in appendix A following the same spirit as in the
work of Viljas {\sl et al.} \cite{Viljas05}. In this appendix we show that the
Fisher-Lee relation is still valid in the general case, {\sl i.e.}: 

\begin{equation}
 T(E) = \text{Tr}\big(\bm{\Gamma}_{L}(E) \bm{G}_{C}(E) \bm{\Gamma}_{R}(E) \bm{G}_{C}^{\dag}(E)\big),
 \label{transmission}
\end{equation}

\noindent
where $\bm{G}_{C}(E)$ is the retarded Green function of the central part defined by the relation:

\begin{equation}
\begin{split}
&\left( \begin{array}{ccc}
E^{+}\bm{S}_{L}-\bm{H}_{L}                 & E^{+}\bm{S}_{LC}-\bm{H}_{LC}                 & 0 \\
E^{+}\bm{S}_{LC}^{\dag}-\bm{H}_{LC}^{\dag} & E^{+}\bm{S}_{C}-\bm{H}_{C}                   & E^{+}\bm{S}_{CR}-\bm{H}_{CR} \\
0                                          &  E^{+}\bm{S}_{CR}^{\dag}-\bm{H}_{CR}^{\dag}  & E^{+}\bm{S}_{R}- \bm{H}_{R}               \\
\end{array}\right) 
\\
& \times \left( \begin{array}{ccc}
\bm{G}_{L}     & \bm{G}_{LC}    & \bm{G}_{LR}        \\
\bm{G}_{CL}    & \bm{G}_{C}     & \bm{G}_{CR}        \\
\bm{G}_{RL}    & \bm{G}_{RC}    & \bm{G}_{R}         \\
\end{array}\right) 
=\left( \begin{array}{ccc}
\bm{I}_{L}     & 0    & 0        \\
0     & \bm{I}_{C}    & 0        \\
0     & 0    & \bm{I}_{R}        \\
\end{array}\right) 
\end{split}
\label{green}
\end{equation}

\noindent where $\bm{I}_{\alpha}$ is the identity matrix and $E^{+}=\lim_{\eta \rightarrow 0}E+i\eta$.
In practice we add a small but finite imaginary part $\eta$ to the energy.
In the following the superscript  $+$ will be omitted for convenience.

Solving Eq. \ref{green} for $\bm{G}_{C}(E)$ yields:

\begin{eqnarray}
\bm{G}_{C}(E)&=&\big(E\bm{S}_{C}-\bm{H}_{C}-\bm{\Sigma}_{L}(E)-\bm{\Sigma}_{R}(E)\big)^{-1} \label{greenc} \\
\bm{\Gamma}_{\alpha}(E)&=& i(\bm{\Sigma}_{\alpha}(E)-\bm{\Sigma}_{\alpha}^{\dag}(E)) \label{gamma}
\end{eqnarray}

 \noindent
$\bm{\Sigma}_{L}$ and $\bm{\Sigma}_{R}$ are the self energy terms which account for the
 coupling of the central part to the leads, {\sl i.e.}:
 
 \begin{equation}
\bm{\Sigma}_{\alpha}(E)=(E\bm{S}_{\alpha C}^{\dag}-\bm{V}_{\alpha C}^{\dag}) \bm{g}_{\alpha}^{S}(E)
(E\bm{S}_{\alpha C}-\bm{V}_{\alpha C})
\label{selfenergy}
\end{equation}

\noindent where $\bm{g}_{\alpha}^{S}(E)$ is the surface Green function of the
uncoupled lead $\alpha$. This surface Green function is calculated in the
absence of the scattering region and thus is different from the Green function 
$\bm{G}_{\alpha}$ defined in Eq.
\ref{green}. The functions $\bm{g}_{\alpha}^{S}(E)$ need to be calculated only 
on the atoms of the leads which
are in contact with the scattering region.
The calculation of the central Green function $\bm{G}_{C}$ is now reduced to the
inversion of a matrix of size $N_{C}\times N_{C}$.

Finally, experiments being performed at small bias voltage and low temperature \cite{Viret06}
Eq.\ref{eq:Itot2} leads to a conductance given by

\begin{equation}
\mathcal{G}=\frac{e^2}{h}T(E_F)
\label{landauer}
\end{equation}

\noindent where $E_F$ is the Fermi level of the system.
Note that the quantum of conductance is here given by $e^{2}/h$ and not $2e^2/h$
because we are considering magnetic materials.

\subsection{Computational details \label{sec:computing}}

The atomic structure of the leads is that of a semi-infinite crystal with
two dimensional periodicity but, due to the presence of the central region, their
electronic structure and magnetic moments are modified
near the scatterer and no longer periodic. However the choice of the scattering region is arbitrary and
can include a part of the leads containing the atomic planes around which the matrix
elements of the Hamiltonian are significantly different from the bulk ones.
We therefore define an effective scattering region by adding a part of the leads
to the central part. In practice for bcc $(001)$ three atomic planes are added on the left-hand side 
of the scatterer and two on the right-hand side. 
The computing process runs as follows.
First we calculate the electronic structure and magnetic moments of the
effective scattering region, and since in most of the cases considered in this work the
left and right leads are identical, we have added periodic boundary conditions
in the three directions. With that procedure the bulk like behavior in the
additional atomic planes is recovered faster. 
We thus obtain the Hamiltonian of the central region $\bm{H}_{C}$ from the 
self-consistent calculation in the effective scattering region, the Fermi level 
being fixed to that of the leads.

\begin{figure}[!fht]
\begin{center}
\includegraphics*[width=\linewidth,angle=0]{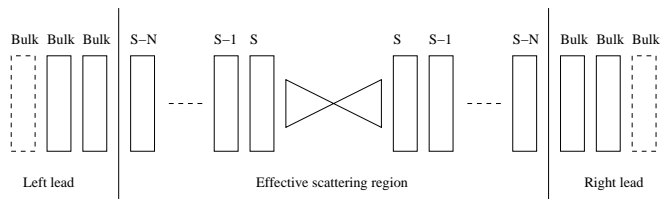}
\end{center}
\caption{Definition of the scattering region for practical calculations. }
\label{fig:system2}
\end{figure}
    
In this geometrical configuration $\bm{g}_{\alpha}^{S}(E)$ (in Eq.\ref{selfenergy}) is the surface Green
function of a truncated bulk in which the layer $\bm{H}_0$  and interlayer $\bm{H}_1$ 
matrix elements (See Appendix B Eq. \ref{HL} ) of the Hamiltonian are the same as in the bulk, 
{\sl i.e.}, they are not modified by surface self-consistency effects. The Hamiltonians 
of the leads $\bm{H}_{L}$ and $\bm{H}_{R}$ are obtained from a bulk calculation.
We then calculate the transmission $T(E)$  (Eq. \ref{transmission}).

\section{Results}

\subsection{A test case: the monatomic iron wire \label{sec:defect}}

We have first considered transport in the simple test case of a monatomic
wire along the $z$ axis where the results can easily be interpreted.
Moreover, this case is relevant for the study of break junctions. 
Indeed experiments on gold or platinum have shown that a short monatomic wire may appear
and stabilize during the breaking process \cite{Smit01,Smit03}.
In iron junctions, both theory and experiments do note show evidence
for the appearance of a  wire \cite{Autes08}.
Nevertheless the contact is still atomic and the geometric environment of the
contact atom is very close to the environment of an atom in a monatomic wire.

In the following we study the influence of various defects on the transmission
probability $T(E)$.

\subsubsection{Geometric defect \label{Sec:geom_defect}}

\begin{figure}[!fht]
\begin{center}
\includegraphics*[width=\linewidth,angle=0]{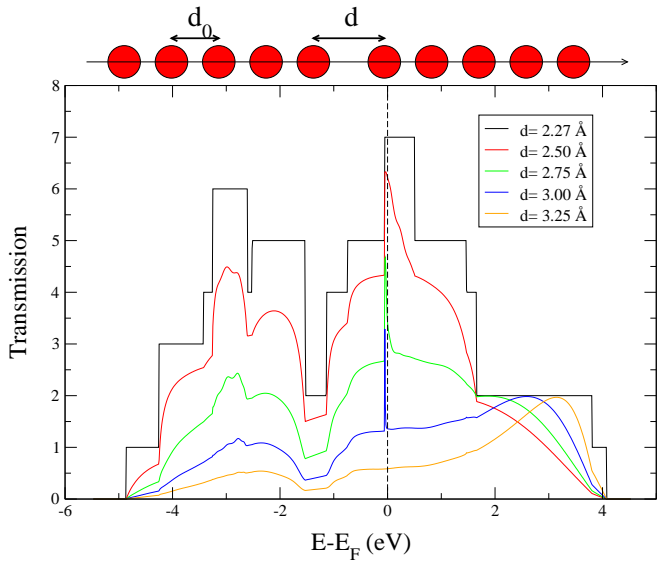}
\end{center}
\caption{ Transmission $T(E)$ for an iron monatomic wire with the geometric defect 
shown in the upper part ($d_0=2.27$\AA).}
\label{fig:geometric}
\end{figure}

First we study the effect of a geometric defect on the transport properties by
considering a perfect infinite wire in which the distance between two atoms is
stretched and  SOC is neglected.
The leads are two semi-infinite wires with an interatomic distance $d_0=2.27 \AA$ 
and magnetic moment $m_0= 3.21 \mu_B$ at equilibrium.
The effective scattering region is made of $10$ atoms at the same equilibrium
 distance, save for  the distance $d$ between the $5th$ and the $6th$ atoms which is stretched from 
equilibrium to $3.25 \AA$ (Fig. \ref{fig:geometric}).
The magnetic moments in the scattering region are computed self-consistently.
Our calculation showed that the magnetic moments of the atoms near the defect
increase with the distance $d$ as expected. 
 
The electronic transmission for different stretchings is reported in Fig. \ref{fig:geometric}.
For the perfect wire (black curve), each channel has a transmission probability
equal to unity  and the total transmission is simply given by the number of electronic
states at energy $E$. As the wire is stretched, the transmission of each channel decreases.

We can see that the reduction of transmission probability is strongly energy dependent.  
The channels corresponding to $d_{xy}$ and $d_{x^2-y^2}$ orbitals ($\delta$ bands), 
which are localized in the plane perpendicular to the transmission direction $z$, 
are more affected by the stretching of the central atom than the channels corresponding to
$s$ and $d_{z^2}$ orbitals. This explains the flattening of the high transmission peaks
at $-3eV$ and $0.25 eV$ which correspond to $\delta$ orbitals.

\subsubsection{Magnetic defect}

We now look at the effect of a domain wall on the electron transport in the
absence of SOC.

\begin{figure}[!fht]
\begin{center}
\includegraphics*[width=\linewidth,angle=0]{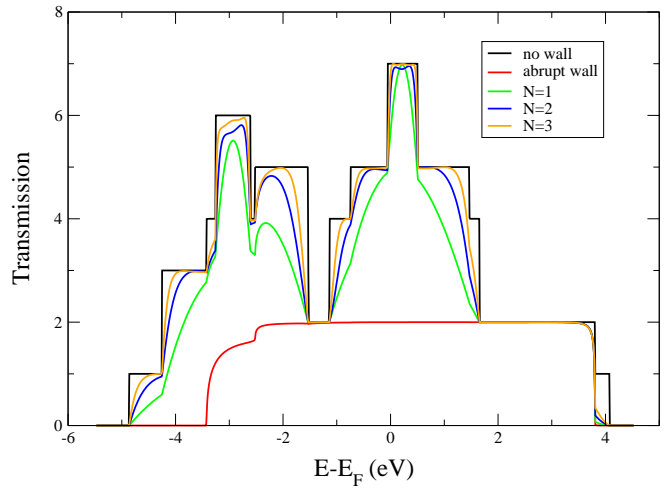}
\end{center}
\caption{ Transmission $T(E)$ of an iron monatomic wire in the presence 
of a $N$ atom long magnetic wall (see text). }
\label{fig:wall}
\end{figure}

The leads are two semi-infinite wires with opposite magnetization directions.
The scattering region contains a magnetic wall of length $N$. The angle 
between the magnetization directions of two consecutive atoms in the wall
is chosen to be constant and equal to $\pi/(N+1)$.

A first calculation shows that the presence of an abrupt wall ($N=0$) results in 
an important loss of transmission. The transmission $T(E)=2$ is due to the $s$ orbitals 
which are the only channels able to transmit in this case.  
Indeed in transition metals, the magnetic moment is mostly carried by the $d$ 
electrons and in an iron wire the $d$ orbitals are totally polarized. Thus no transmission 
is possible through the wall. On the other hand, $s$ orbitals are only
partially polarized and are able to transmit.
 
When the magnetization is flipped by $\pi/2$ on one atom between the leads,
leading to a domain wall with $N=1$, the $d$ channels open and the 
transmission increases. This spin-flip transmission is made possible by the states 
in the flipped atom that mix spin up and spin down orbitals.  
As the length of the wall increases, the transmission quickly tends towards 
the transmission of a perfect wire with no domain wall.

Thus, in the ballistic regime, a magnetic wall a few atom wide  
has almost no effect on the electron transmission. This results is in good agreement 
with the calculation of Velev and Butler \cite{Velev04} who showed that large magneto-resistance
is expected only in contacts with very narrow domain walls. 

\subsubsection{Influence of spin orbit-coupling \label{sec:amr-wire}}

\begin{figure}[h]
\begin{center}
\includegraphics*[width=\linewidth,angle=0]{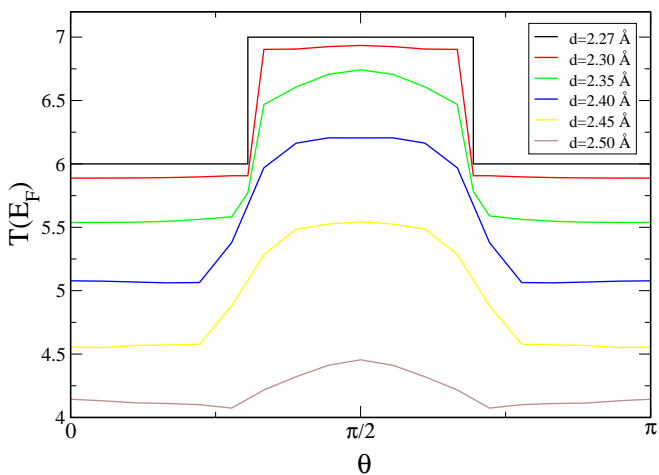}
\end{center}
\caption{ Transmission at the Fermi energy $T(E_F)$ in an iron monatomic wire with the
geometric defect shown in Fig. \ref{fig:geometric}, as a function 
 of the magnetization direction $\theta$. }
\label{fig:amr-wire}
\end{figure}

In this section SOC is taken into account. Therefore the electronic structure depends on the
 magnetization direction and  anisotropic effects appear.
In a perfect monatomic wire of a transition metal, SOC removes the degeneracy of the $\delta$
bands except when 
the magnetization is perpendicular to the wire ($\theta=\pi/2$) \cite{Autes06}.
 In iron, these bands lie around the 
Fermi level and the splitting can be strong enough to bring one of the bands above it when $\theta$ 
varies. Thus, the conductivity of the wire 
(which is proportional to the number of bands crossing its Fermi level) decreases by $1$ as 
the magnetization orientation is switched from $\pi/2$ to $0$ (Fig. \ref{fig:amr-wire}, black curve).    
This anisotropic magneto-resistance (AMR) in $3d$ transition metal wires has already been
reported \cite{Velev05,Viret06}. Note that here and in all the following calculations in which
 $\theta$ is varied, a penalization function has been added to the Hamiltonian in order 
 to keep a fixed value of $\theta$ during the self-consistent
iteration process (See Ref. \cite{Autes06}).

If a geometric defect is present in the wire, the channel transmission is no more equal to unity. 
 The curve of the transmission as a function of the magnetization direction for the
 same type of defect as in Sect.\ref{Sec:geom_defect} is no longer a step 
 between two integer values but becomes continuous with a maximum reached at $\theta=\pi/2$ 
 (Fig. \ref{fig:amr-wire}). The AMR is softened by the presence of the defect as already
  reported by Jacob et al. \cite{Jacob}.
We have seen in Sec. \ref{Sec:geom_defect} that the $\delta$ channel was  the most affected 
by a geometric defect in a wire. Since the AMR in an iron  wire is only due to electrons in
 these $\delta$ bands it is strongly affected by the presence of a defect 
 (Note that the domain of stretching is different in Fig. \ref{fig:amr-wire} and in Fig. \ref{fig:geometric}).

\subsection{Influence of the atomic contacts}

We have seen in  Sec. \ref{sec:defect} that 
the transmission of an iron atomic wire is affected by geometric or magnetic defects. 
Another strong effect arises since in practice the wire is always connected to some leads.
These contact effects can become very important for instance in break junction experiments where
 the leads are bulk-like while the scattering region is quasi one-dimensional.
To have a better understanding of the role of the contacts in the Landauer formalism 
applied to break junctions, we studied
the transmission of some simple systems.

\subsubsection{The contact effect in a $s$ band model}

 \begin{figure}[!fht]
 \begin{center}
\includegraphics*[width=0.8\linewidth,angle=0]{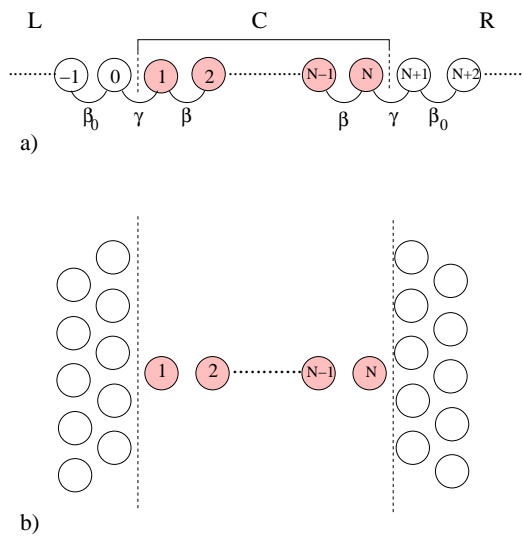}

\end{center}
 \caption{ Geometries used in our calculations. a) N atom long wire with hopping integral 
 $\beta$ connected to two semi-infinite wires with hopping integral $\beta_0$ by means of the contact 
hopping integral $\gamma$. b) N atom long wire between two fcc $(001)$ surfaces. The atoms
$1$ and $N$ are in four-fold position relative to the surface. The distance between atoms
  in the wire, in the leads and at the contact is given by the bulk interatomic distance at
   equilibrium of the considered material. }
 \label{fig:geom}
 \end{figure}

The simplest way to model the contact is to consider a non-magnetic wire of $N$ atoms with a single $s$
orbital per site and an hopping integral $\beta$ between first nearest neighbours. This wire is connected
to two semi-infinite wires which have a different hopping integral $\beta_0$ but the same atomic level.
The contact is established by a hopping integral $\gamma$ between the leads and the edge of the
finite wire (Fig. \ref{fig:geom} a). The three hopping integrals are taken negative, overlap integrals
are neglected, and all first neighbour interatomic  distances are equal to unity.  
The transmission of this system can be calculated analytically (see Appendix \ref{sec:finite_wire}) 
and is given by:

 \begin{figure}[th]
 \begin{center}
\includegraphics*[width=\linewidth,angle=0]{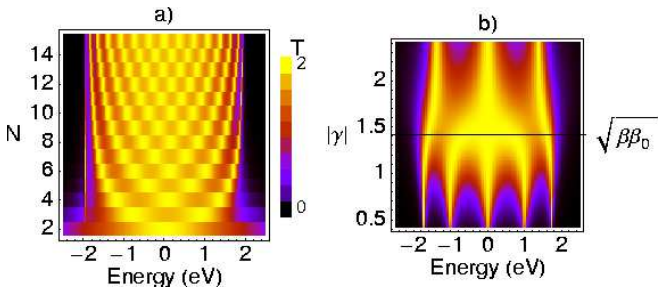}
\end{center}
 \caption{Transmission of the system shown in Fig. \ref{fig:geom}. a) calculated from Eq. \ref{eq:T_N} 
 as a function of energy and of the number $N$ of atoms in the central wire, 
  with $\beta_0=-2$, $\beta=-1$ and $\gamma=-1.2$. b) as a function of energy and of the strength
  $|\gamma|$ of the contact, with $\beta_0=-2$, $\beta=-1$ and $N=5$.}
 \label{fig:T_N}
 \end{figure}

\begin{equation}
\begin{split}
& T_N(E) =      \\
&2 \biggl \lvert 
        \frac{2 \sin k \sin k_0}
             {\textstyle \frac{\gamma^2 e^{ik_0}}{\beta \beta_0} \sin (N-1)k - 2  \sin Nk  + \frac{\beta \beta_0}{\gamma^2 e^{ik_0}} \sin (N+1)k  } 
 \biggr \rvert^2
\end{split}
\raisetag{52pt}
\label{eq:T_N}
\end{equation}

\noindent
where $T_N(E)$ is the transmission coefficient of a $N$ atom long wire at energy $E$ with
  $E=2\beta \cos k=2\beta_0 \cos k_0 $.

In the following we study the case $|\beta| < |\beta_0|$, {\sl i.e.}, the dispersion of the
central part is narrower than that of the leads.
In Fig. \ref{fig:T_N}a we plot the transmission $T_N(E)$ when $|\beta|<|\gamma|<|\beta_0|$,
for different values of $N$. Note that $k$ is imaginary when $E>2|\beta|$.
In the limit of large $N$ the transmission is non-vanishing only inside the energy band of
the $N$ atom wire ( $-2|\beta|\leq E \leq 2|\beta|$ ). Inside this band, the transmission oscillates
and presents a number of peaks that increases with $N$ (there are $N$ peaks for a wire of $N$ atoms
for this specific choice of $\beta$, $\beta_0$ and $\gamma$). It is interesting to note that, 
in the middle of the band (where $E=0$ and $k=k_0=\pi/2$), the transmission oscillates between 
two values when $N$ goes from odd to even. If $N$ is odd, $T_N(0)=2$ and if $N$ is even 
$T_N(0)=8(\gamma^2/(\beta\beta_0)+(\beta\beta_0)/\gamma^2)^{-2}$. When  $\gamma=-\sqrt{\beta\beta_0}$,
 this oscillation disappears. Finally for $N$ small ($N\leq 5$) the transmission has a significant 
 exponentially decaying tail outside the energy range  $[-2|\beta|, 2|\beta|]$.

In Fig. \ref{fig:T_N}b, we plot the $\gamma$ dependence of the transmission of a five atom wire.
For $|\gamma|=\sqrt{\beta\beta_0}$, the transmission is very close to $2$ on the whole energy spectrum.
For small values of $|\gamma|$, $N$ sharp peaks appear at energies close to the eigenvalues
of the isolated finite chain. When $|\gamma|$ increases the peaks broaden and are shifted so that
some of them disappear.

Thus, the number of oscillations in the transmission curve increase with the number of 
atoms in the wire  and the strength of the contact $\gamma$ controls the amplitude of 
these oscillations.
These results show that  a too strong or too weak  contact kills the transmission while with
an appropriate choice of $\gamma$ the transmission can be perfect.

\begin{figure}[th]
 \begin{center}
\includegraphics*[width=0.8\linewidth,angle=0]{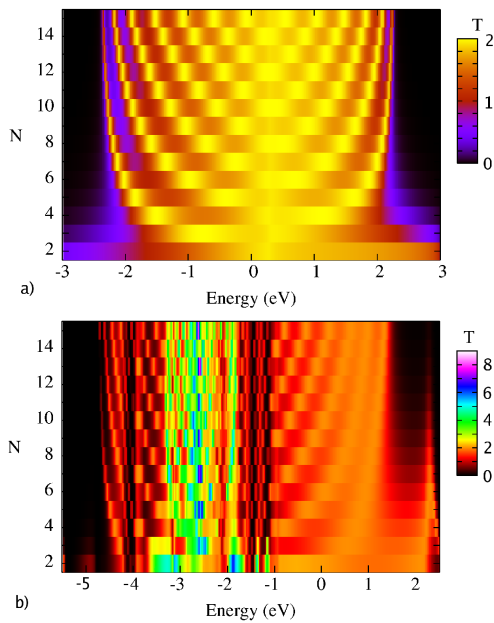}
\end{center}
 \caption{Transmission $T(E)$ of a N atom Au wire between two Au$(001)$ surfaces. 
 a) $s$ parametrization. b) $spd$ parametrization. Note the complex behaviour in the $d$ band
 energy range. $E_F$ is the zero of energy.}
 \label{fig:s}
 \end{figure}

Let us now consider a more complex geometry where the leads are two semi-infinite surfaces as shown in
Fig. \ref{fig:geom}b. We illustrate this case for gold using first a basis set limited to overlapping $s$
orbitals
with a parametrization taken from Mehl and Papaconstantopoulos \cite{Mehl96}. Note that however we assume
that the onsite term does not vary with the atomic environment and is chosen as the zero of  energies. This
model can be used as a first approach since the electronic conduction in gold is dominated by the
$s$ electrons as $E_F$ belongs to the $s$ band.
The choice of gold is especially interesting because experiments on gold atomic contacts
have shown its ability to form monatomic wires several atom long. Furthermore the conductivity of
these wires exhibits specific features such as quantization and parity oscillations \cite{Agrait03}. 
The inter-atomic distance in the wire and in the contacts is the nearest neighbour
distance in bulk gold at equilibrium ($2.88 \AA$), and the surfaces of the two leads are fcc$(001)$.
The results are shown in Fig. \ref{fig:s}a. 
As for the perfect infinite wire we observe that the transmission curves oscillate as a function 
of $N$ and $E$.
In the middle of the band, the transmission is close to $2$ and exhibits even-odd oscillations. 

Additional calculations with different distances between the finite wire and the surfaces revealed that,
 as in the infinite wire case, the strength of the contact affects the amplitude of the oscillations.  
From these results, we can conclude that in the case of a pure $s$ band, the choice of wire
 leads or bulk-like leads does not much modify the transmission of the system. 

\subsubsection{The contact effect in metals with $spd$ bands\label{sec:contacts-spd}}

To investigate the contact effect on a system with a realistic electronic structure we have calculated
the transmission for a gold wire of $N$ atoms connecting two $(001)$ gold surfaces using the $spd$
TB parametrization of Mehl and Papaconstantopoulos \cite{Mehl96}, and neglecting SOC.
The results are shown in Fig. \ref{fig:s}b.
We can see that in the energy range where only the $s$ band is present, the transmission curves are similar
 to Fig. \ref{fig:s}a, with oscillations as a function of $N$ and $E$ and a transmission close to $2$.
  In the energy range corresponding to $d$ states $T(E)$ has a complex behaviour which
does not show simple oscillations. It is clear that the presence of contacts 
  strongly affects the transport of $d$ electrons.
  However in gold the $d$ electrons do not participate in the conductance
  at low bias voltage since the Fermi level has a pure $s$ character. Thus a gold atomic contact exhibits a 
quantization of its conductance and odd-even oscillations as observed experimentally \cite{Smit03} 
and explained theoretically \cite{Vega04}.  

\begin{figure}[!fht]
 \begin{center}
\includegraphics*[width=\linewidth,angle=0]{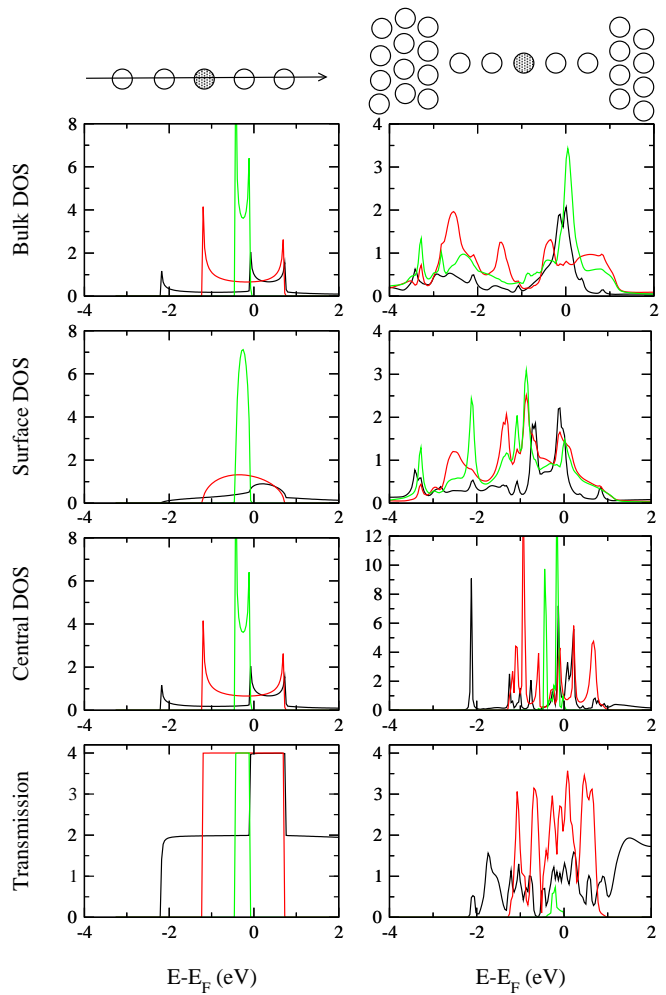}
\end{center}
 \caption{Bulk (a),  local DOS at the surface of the leads (b) and  on the central atom (c) of the wire, and 
 transmission (d) in an infinite Fe wire (left) or in a 5 atom Fe wire connecting two bcc$(001)$Fe surfaces (right).
  The contributions of the $\sigma$ (black), $\pi$ (red) and $\delta$ (green) channels (see text) have been
  resolved. }
 \label{fig:mode_dec}
 \end{figure}

On the opposite, in transition metals like iron, $s$, $p$ and  $d$ electrons are
 present at the Fermi level and the
 behaviour of the atomic contact conductivity is more intricate \cite{Smogunov06}. 
To illustrate qualitatively the importance of contacts in such a case we have calculated 
the transmission of a non-magnetic iron wire of 5 atoms connecting two bcc $(001)$ leads and decomposed it
into the contributions of the different transmission channels. The results are shown in
 Fig. \ref{fig:mode_dec}. 
 We have also reported the local density of states (LDOS) on the central atom of
  the wire, on the surface of the leads when disconnected from the wire derived from
  $g_{\alpha}^S(E)$ as explained in Sec. \ref{sec:computing}, and in the bulk.
 For the sake of comparison we have calculated the same 
  quantities in the case of a perfect infinite wire. 

The $\sigma$, $\pi$ and $\delta$ decomposition corresponds to the three band symmetries found in the 
perfect wire and defines its transmission channels.  The wire being along the $z$ axis, the $\sigma$ channel 
 results from the hybridization of
$s$,$p_z$ and $d_{z^2}$ orbitals, the $\pi$ channel from $p_x$, $p_y$, $d_{xz}$ and $d_{yz}$,
and the $\delta$ channel from $d_{x^2-y^2}$ and $d_{xy}$ orbitals.

 In the case of the wire between two surfaces $\sigma$, $\pi$ and $\delta$ orbitals become coupled.
 Accordingly the transmission channels have no longer a pure $\sigma$, $\pi$ or $\delta$ character.
 The general way of determining the transmission channels is to calculate the $t$ matrix defined
 as
 
 \begin{equation}
 t(E)=(\Gamma_L)^{1/2}G_C(\Gamma_R)^{1/2}
 \end{equation}
 
 \noindent
 such that $T(E)=\text{Tr}(tt^{\dag})$. The transmission channels are the eigenvectors of 
 $tt^{\dag}$ and the corresponding eigenvalues are the associated transmission factors \cite{Cuevas98,Jacob06}. Since the choice
 of the scattering region $C$ is somewhat arbitrary we took the central atom of the finite wire.
 The calculation shows that for  a wire as short as 5 atoms the transmission channels have nearly a
 pure $\sigma$, $\pi$ or $\delta$ character. However the value of the transmission factor is no
 longer quantized but presents sharp oscillations as a function of $E$ (See Fig. \ref{fig:mode_dec}).
 In order to obtain a physical insight into this result we also show in Fig. \ref{fig:mode_dec}
  the sum of the LDOS projected on the various orbitals
 contributing to each channel. For the central atom of the wire the LDOS have some similarities
 with the ones of the perfect wire, especially the $\delta$ LDOS. Indeed all the weight of the
 $\delta$ LDOS is concentrated in the same energy range as in the perfect wire. This is no
 longer true for the LDOS of the leads, since the symmetries are completely changed. As a result
 the various LDOS have a significant weight on the total energy spectrum and are thus lowered. 
Consequently the transmission channels are strongly affected. In particular the contribution of
the $\delta$ channel is almost zero.

 Finally let us note that the connection between the lead and the wire considered here
 is very abrupt and a possible origin of the weak transmission could be this unrealistic
 geometry. However we do not believe in the pertinence of this suggestion  since we have seen
 that for a pure $s$ band this geometry can lead to almost perfect transmission.
 Moreover similar calculations for magnetic Fe in an $spd$ model, where we assumed a smoother contact
 (small pyramids), gave also low channel transmission \cite{Autes08}. 

\subsubsection{Anisotropic magnetoresistance effects in Fe atomic contacts}

We have seen in Sec.\ref{sec:amr-wire} that an infinite iron monatomic wire shows a step-like
variation of $T(E_F)$ as a function of the magnetization direction (Fig. \ref{fig:amr-wire}).
This abrupt AMR takes its origin in the behaviour of the $\delta$ band and
is softened by the stretching of one bond. In the previous Sec. 
(\ref{sec:contacts-spd}) we have also demonstrated the influence of contacts 
on the transmission probability of each channel which may almost vanish. 
It is therefore interesting to investigate the role of contacts on the AMR.

\begin{figure}[!fht]
  \begin{center}
    \includegraphics*[width=\linewidth,angle=0]{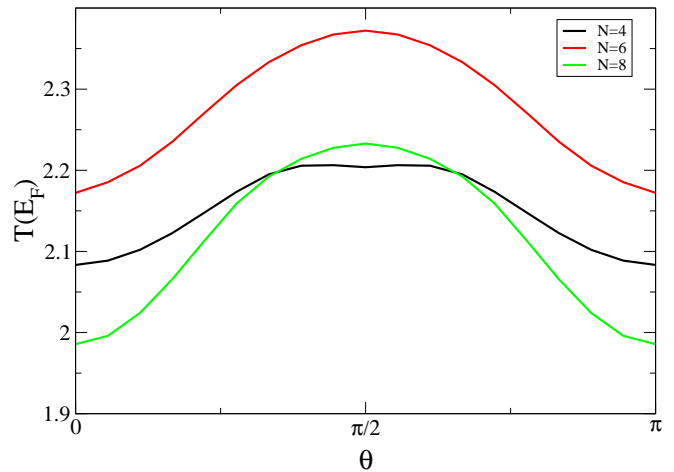}
  \end{center}
  \caption{Transmission at the Fermi energy of a $N$ atom long Fe wire connecting two
  bcc $(001)$Fe surfaces as a function of the magnetization direction. }
  \label{fig:amr-wire_surf}
\end{figure}

We have calculated the transmission at the Fermi energy in an iron wire  of $N$ atoms connected
to two bcc$(001)$ Fe leads for different magnetization directions. 
The results are shown in Fig.\ref{fig:amr-wire_surf} as a function of $N$.

The AMR effect is still present, but the curves are continuous and the difference of transmission
 for magnetizations along or perpendicular to the wire is lower than $1$.
  The switching between two quantized conductance values predicted in the perfect infinite wire 
  (Fig. \ref{fig:amr-wire}, black curve) disappears.

\section{Conclusion}

The idealized model of an infinite monatomic wire has often been used
to interpret the conductance measurements in atomic contacts. However, even though a wire
may exist between two electrodes, it may present some defects. Furthermore this wire is necessarily
finite and the contact between the two leads cannot be ignored.
In this work we have investigated, on the one hand, the influence of structural (stretching of a bond) and
magnetic (domain wall) defects on the transport properties of an infinite wire. On the other
hand, we have studied the contact effect, both in a $1D$ model and in a more elaborated
geometry in which the leads are represented by semi-infinite $3D$ crystals (See Fig. \ref{fig:geom}).

The role of defects in a Fe monatomic wire has been computed with an elaborate tight-binding
Hamiltonian written in an $spd$ basis set. As expected the stretching of one bond
or the presence of a magnetic domain wall decrease the transmission factor (which is no
longer quantized) the more as the defect is pronounced (large stretching or abrupt wall)
except outside the range of the $d$ bands ({\sl i.e.}, between 2 and 4 eV above the Fermi level).
Surprisingly the effect of a domain wall almost vanishes as soon as its width
overcomes three atoms.
When spin-orbit coupling is taken into account, the step-like behaviour of $T(E_F)$ as 
a function of the magnetization direction $\theta$ ({\sl i.e.}, the AMR effect) is softened
by the stretching of one bond in a perfect infinite wire.

The influence of contacts has been compared for two metals: Au and Fe in which the
electronic states at the Fermi level have $s$ and $d$ character, respectively.
In gold the effect of contacts on the transmission factor is strong in the
energy range corresponding to $d$ bands, but relatively limited around the
Fermi level. Note however in the latter case the presence of smooth odd-even oscillations.

In iron the transmission factor presents very sharp oscillations as a function of
energy in the presence of contacts, and some transmission channels can even
almost disappear, in particular the channel of $d_{xy}$ and $d_{x^2-y^2}$ character ($\delta$ band).
This latter result casts some serious doubts on the interpretation of the AMR effects in magnetic
contacts based on the band structure of a perfect infinite wire.
In conclusion we have demonstrated that contact effects strongly change the electronic transmission
of a monatomic wire in transition metals. Thus it is hoped that by modifying the contacts
one could tailor the transport properties of nanowires.

\acknowledgments

\appendix

\section{Derivation of the Fisher Lee formula in the general case}
The aim of this appendix is to prove the Eq.\ref{transmission} of the main text which
gives the electronic transmission in the general case where the basis 
set is non-orthogonal and not necessarily real and in which the matrix
elements of the hamiltonian may be complex. 

We consider an electron which at time t=$-\infty$ is described by an incident Bloch wave
$\psi^i_{(n)}$ of the left lead, i.e., obeying:

\begin{equation}
({\bf H}_L-E_{(n)}{\bf S}_L)\psi^i_{(n)}=0
\label{eq:SchL}
\end{equation}

\noindent and switch on adiabatically the connection to the scattering
region and the right lead. At a time t$\leq 0$ the hamiltonian is thus
${\bf H}(t)= {\bf H}_L+({\bf H}_C+{\bf H}_R)\exp(\epsilon t)$, $\epsilon$ 
is a small positive factor which ensures that the connection is established
adiabatically and $\psi({\bf r},t)$ can be expanded as a linear
combination of atomic spin-orbitals with time dependent coefficients
$a_{(n)\alpha}(t)$. At t=0 the wave function $\psi({\bf r},t)$ has evolved towards
a stationary state of ${\bf H}$ with the same energy as the incident wave that
will be denoted in the following as $\psi$ for short.

The charge $N_{\Omega}$
contained in a given part $\Omega$ of the system ($\Omega= L$(left lead),
$C$(scattering region), $R$(right lead), $L+C$, $C+R$) at time $t$ is defined
according to the Mulliken population analysis and is written as (after having
summed over all possible incident waves):

\begin{eqnarray}
N_{\Omega}(t)&=& Re[\sum_{n_{\text{occ}},\alpha_{\Omega},\beta}a_{(n)\alpha_{\Omega}}(t)a_{(n)
\beta}^*(t)S_{\beta\alpha_{\Omega}}] \nonumber \\ &=& Re\, \text{Tr}_{\Omega}\,{\bf P}(t){\bf S}
\label{eq:nomega}
\end{eqnarray}

\noindent where $\alpha_{\Omega}$ is an atomic spin-orbital belonging to the part
$\Omega$ and $\beta$ is any spin-orbital of the system,
${\bf S}$ is the overlap matrix
and $P_{\alpha\beta}(t) = \sum_{n_{\text{occ}}}a_{(n)\alpha}(t)a_{(n)\beta}^*(t)$ are
the elements of ${\bf P}$. Note the presence of the real part in Eq. \ref{eq:nomega} which
is necessary when the hamiltonian or the basis set is complex. In the following the ${\bf P}$ matrix is divided
into blocks, similarly to ${\bf H}$ and ${\bf S}$ (see Eq.\ref{hamiltonian} of the main text).
It is convenient to define another population $N_{\Omega}'$ which is
obtained by retaining in Eq.\ref{eq:nomega} only the overlap populations
between orbitals included in $\Omega$, i.e.: 

\begin{equation}
N_{\Omega}'(t)=\text{Tr}\,{\bf P}_{\Omega}(t){\bf S}_{\Omega}.
\label{eq:npomega}
\end{equation}

The current $I_{L\rightarrow R}$ coming from the left
is given either by the electrons flowing from $L+C$ or by those entering
in the right lead $R$. Thus:

\begin{equation}
I_{L \rightarrow R}=\frac{1}{2} (\frac{\partial N_{L+C}}{\partial t}-
\frac{\partial N_R}{\partial t})=\frac{1}{2} (\frac{\partial N_{L+C}'}{\partial t}-
\frac{\partial N_{R}'}{\partial t})
\label{eq:ILR}
\end{equation}

\noindent since the overlap populations between $L+C$ and $R$ cancel in this difference.
Let us now calculate $\frac{\partial N_{\Omega}'}{\partial t}$:

\begin{eqnarray}
\frac{\partial N_{\Omega}'}{\partial t}&=&\frac{\partial}{\partial t} [\sum_{n_{\text{occ}}, \alpha_{\Omega}}
a_{(n)\alpha_{\Omega}}(t)a_{(n)\alpha_{\Omega}}^*(t)  \label{eq:DNPDT} 
\\ &+&  Re  \sum_{n_{\text{occ}}, \alpha_{\Omega},
\beta_{\Omega}\neq \alpha_{\Omega}}a_{(n)\alpha_{\Omega}}(t)a^*_{(n)\beta_{\Omega}}(t) S_{\beta_{\Omega}\alpha_{\Omega}}] \nonumber 
\end{eqnarray} 

From the time dependent Schroedinger equation written in the non-orthogonal basis set we have:

\begin{eqnarray}
\frac{\partial a_{(n)\alpha_{\Omega}}(t)}{\partial t}&=& \frac{1}{i\hbar} [H_{\alpha_{\Omega}\alpha_{\Omega}}(t)a_{(n)\alpha_{\Omega}}(t)  \label{eq:DADT} \\ &+&
\sum_{\beta \neq \alpha_{\Omega}} [H_{\alpha_{\Omega} \beta}(t)-i\hbar S_{\alpha_{\Omega} \beta}\frac{\partial}{\partial t}]a_{(n)\beta}(t)]. \nonumber
\end{eqnarray}

Substituting Eq.\ref{eq:DADT} for $\frac{\partial a_{(n)\alpha_{\Omega}}}{\partial t}$ into
the first term of the right-hand side of Eq.\ref{eq:DNPDT} yields:
   
\begin{eqnarray}
\frac{\partial N_{\Omega}'}{\partial t}&=&\frac{1}{i\hbar}\sum_{n_{\text{occ}}, \alpha_{\Omega}, \beta_{\overline{\Omega}}}a_{(n)\alpha_{\Omega}}^*(t)
[H_{\alpha_{\Omega} \beta_{\overline{\Omega}}}(t) \nonumber \\ 
&-&i\hbar S_{\alpha_{\Omega} \beta_{\overline{\Omega}}}\frac{\partial}{\partial t}]a_{(n)\beta_{\overline{\Omega}}}(t)
+c.c. \label{eq:DNPDT2}
\end{eqnarray}

\noindent $\Omega+\overline{\Omega}$ denoting the whole system. At $t=0$ the state $n$ is a stationary state of energy $E_{(n)}$,
thus $\partial a_{(n)\alpha}/\partial t$ can be replaced by $-\frac{i}{\hbar}E_{(n)}a_{(n)\alpha}(0)$. Finally
substituting Eq.\ref{eq:DNPDT2} for $\partial N'_{\Omega}/\partial t$ in Eq.\ref{eq:ILR} we find:

\begin{equation}
I_{L\rightarrow R}=\frac{1}{i\hbar} \text{Tr}_C\, {\bf W}_{CR}(E_{(n)}){\bf P}_{RC} + c.c.
\label{ILR2}
\end{equation}

\noindent with:

\begin{equation}
{\bf W}(E_{(n)})={\bf H}-E_{(n)}{\bf S}.
\label{eq:W}
\end{equation}

The response of the system to $\psi^i_{(n)}$ is denoted as $\psi'$, $\psi=\psi^i_{(n)}+\psi'$ obeying
the Schroedinger equation of the whole system:

\begin{equation}
({\bf H}-E_{(n)}{\bf S})(\psi^i_{(n)}+\psi')=0.
\label{eq:SFS}
\end{equation}

In the non-orthogonal basis set the vectors $\psi^i_{(n)}$ and $\psi'$ can be decomposed into blocks
corresponding to the three parts of the system ($L, C, R$), i.e., 

\[ \psi^i_{(n)}= 
\left(\begin{array}{c}
{\bf a}^i_{(n)L} \\
{\bf 0}          \\
{\bf 0}  \end{array}  \right)
\quad \text{and} \quad 
 \psi=\left(
 \begin{array}{c}
 {\bf a}_{(n)L} \\
  {\bf a}_{(n)C} \\ {\bf a}_{(n)R}
   \end{array} \right) \]

\noindent
respectively. Substituting these
components into Eq.\ref{eq:SFS} with ${\bf H}_{LR}={\bf S}_{LR}=0$ and taking Eq.\ref{eq:SchL} 
into account yields after simple algebraic manipulations:

\begin{equation}
{\bf a}_{(n)C}={\bf G}_C(E_{(n)}){\bf W}_{CL}(E_{(n)}){\bf a}^i_{(n)L} 
\label{eq:aC}
\end{equation}
\begin{equation}
{\bf a}_{(n)R}={\bf G}_{RC}(E_{(n)}){\bf W}_{CL}(E_{(n)}){\bf a}^i_{(n)L} 
\label{eq:aR}
\end{equation}

\noindent with:

\begin{equation}
{\bf G}(E_{(n)})=(E_{(n)}{\bf S}-{\bf H})^{-1}.
\label{eq:G}
\end{equation}

\noindent Finally from Eq.\ref{green} of the main text it is easy to show that:

\begin{equation}
{\bf G}_{RC}(E_{(n)})={\bf g}_R(E_{(n)}){\bf W}_{RC}(E_{(n)}){\bf G}_C(E_{(n)})
\label{eq:GRC}
\end{equation}

\noindent with:

\begin{equation}
{\bf g}_R(E_{(n)})=(E_{(n)}{\bf S}_R-{\bf H}_R)^{-1}.
\label{eq:gR}
\end{equation}

\noindent Substituting Eq.\ref{eq:GRC} for ${\bf G}_{RC}$ into Eq.\ref{eq:aR} and
using Eqs.\ref{eq:aC} and \ref{eq:aR} to obtain ${\bf P}_{RC}$ leads to:

\begin{eqnarray}                
I_{L\rightarrow R}&=&\frac{1}{i\hbar}\sum_{n_{\text{occ}}}[{\bf a}^{i\dag }_{(n)L}
{\bf W}_{LC}(E_{(n)}){\bf G}^{\dag}_C(E_{(n)}){\bf W}_{CR}(E_{(n)}) \nonumber \\
 &\times& {\bf g}_R(E_{(n)}){\bf W}_{RC}(E_{(n)})
{\bf G}_C(E_{(n)}){\bf W}_{CL}(E_{(n)}){\bf a}^{i}_{(n)L} \nonumber  \\
 &-& c.c.]
\label{eq:ILR3}
\end{eqnarray}

\noindent where ${\bf G}^{\dag}$ is the hermitian conjugate of ${\bf G}$, or:

\begin{eqnarray}                
I_{L\rightarrow R}&=&\frac{1}{i\hbar}\sum_{n_{\text{occ}}}[{\bf a}^{i\dag}_{(n)L}
{\bf W}_{LC}(E_{(n)}){\bf G}^{\dag}_C(E_{(n)}) \nonumber \\ 
&\times &{\bf W}_{CR}(E_{(n)})({\bf g}_R(E_{(n)})-
{\bf g}^{\dag}_R(E_{(n)})){\bf W}_{RC}(E_{(n)}) \nonumber \\
&\times &{\bf G}_C(E_{(n)}){\bf W}_{CL}(E_{(n)}){\bf a}^{i}_{(n)L}].
\label{eq:ILR4}
\end{eqnarray}

\noindent From Eqs.\ref{selfenergy} and \ref{gamma} of the main text we have:

\begin{eqnarray}
{\bf W}_{CR}(E_{(n)})({\bf g}_R(E_{(n)})-
{\bf g}^{\dag}_R(E_{(n)})){\bf W}_{RC}(E_{(n)})= \nonumber \\
-i{\bf \Gamma}_R(E_{(n)})\,\,\,\,\,\,
\label{eq:GammaR}
\end{eqnarray}

\noindent thus:

\begin{eqnarray}
I_{L\rightarrow R}&=&-\frac{1}{\hbar}\sum_n\int_{-\infty}^{E_{F_L}}{\bf a}^{i \dag}_{(n)L}
{\bf W}_{LC}(E){\bf G}^{\dag}_C(E) \nonumber \\ &\times &{\bf \Gamma}_R(E)
{\bf G}_C(E){\bf W}_{CL}(E) \nonumber \\ &\times &
{\bf a}^{i}_{(n)L}\delta(E-E_{(n)})\text{d}E
\label{eq:ILR5}
\end{eqnarray}

\noindent where $E_{F_L}$ is the Fermi level of the left lead.
The integrand can be written as a trace if we introduce the square matrix
of dimension $N_L\times N_L$ with elements $\sum_n a^{i*}_{(n)\alpha_L}a^{i}_{(n)\beta_L}
\delta(E-E_{(n)})$ which is nothing but $i({\bf g}_L(E)-
{\bf g}^{\dag}_L(E))/2\pi$ (with obvious notations) in the general case where ${\bf H}$
and the basis set may be complex. By making use of the cyclic invariance of
the trace we obtain finally:

\begin{eqnarray}
I_{L\rightarrow R}&=&-\frac{1}{\hbar}\int_{-\infty}^{\infty} \text{Tr}\, 
({\bf G}^{\dag}_C(E){\bf \Gamma}_R(E){\bf G}_C(E){\bf \Gamma}_L(E)) \nonumber \\ &\times &
f(E-E_{F_L})\text{d}E
\label{eq:ILR6}
\end{eqnarray}

\noindent where $f$ is the Fermi function.

The current due to an incident wave from the
right lead $I_{R\rightarrow L}$ is obtained by interchanging the $R$ and $L$ indices.

\noindent The total current is

\begin{equation}
I=I_{L\rightarrow R}-I_{R\rightarrow L}
\label{eq:Itot}
\end{equation}

\noindent and, since the current should vanish whatever the system when $E_{F_R}=E_{F_L}$:

\begin{equation}
I =-\frac{1}{\hbar}\int_{-\infty}^{\infty} T(E) \times  (f(E-E_{F_L})-f(E-E_{F_R}))\text{d}E.
\label{eq:Itot2}
\end{equation}

with

\begin{equation}
T(E)= \text{Tr} ({\bf \Gamma}_L(E){\bf G}_C(E){\bf \Gamma}_R(E){\bf G}^{\dag}_C(E) )
\label{eq:T(E)}
\end{equation}

\noindent
$T(E)$ is the electronic transmission
given by the Fisher-Lee relation (Eq.\ref{transmission} of the main text).

\noindent Actually it can be shown that \cite{Datta2}:

\begin{eqnarray}
 \text{Tr}\, 
({\bf \Gamma}_L(E){\bf G}_C(E){\bf \Gamma}_R(E){\bf G}^{\dag}_C(E) )= \nonumber \\
\text{Tr}\, ({\bf \Gamma}_R(E){\bf G}_C(E){\bf \Gamma}_L(E){\bf G}^{\dag}_C(E)).
\end{eqnarray}

\section{Calculation of the surface Green function}

\begin{figure}[!fht]
\begin{center}
\includegraphics*[width=0.5\linewidth,angle=0]{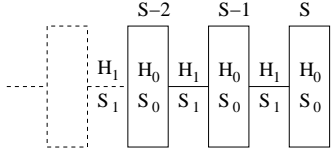}
\end{center}
\caption{ Each lead can be built from a layer periodically repeated. }
\label{fig:sgf}
\end{figure}

To calculate the Green function at the surface of the lead $g_{\alpha}^{S}(E)$ 
(See Sec.\ref{sec:transport} and
Sec. \ref{sec:computing}),
we use an iterative scheme.  Each lead has two dimensional periodicity and can be viewed as a
semi-infinite succession of identical layers (Fig. \ref{fig:sgf}) with
the same interlayer spacing.
The layers contain several atomic planes so that there are interaction only  between consecutive layers.

The Hamiltonian of the semi-infinite lead can thus be written as

\begin{equation}
\bm{H}_{\alpha}=\left(\begin{array}{ccccc}
\ddots & \ddots             & \ddots            & \vdots           & \vdots  \\
\ddots & \bm{H}_{0}         & \bm{H}_{1}        & 0                & 0       \\
\ddots & \bm{H}_{1}^{\dag}     & \bm{H}_{0}        & \bm{H}_{1}       & 0       \\
\ldots & 0                  & \bm{H}_{1}^{\dag}    & \bm{H}_{0}       & \bm{H}_{1}    \\  
\ldots & 0                  & 0                 & \bm{H}_{1}^{\dag}   & \bm{H}_{0}    \\  
\end{array}\right) 
\label{HL}
\end{equation}

\noindent where $\bm{H}_{0}$ is the Hamiltonian of a layer and $\bm{H}_{1}$ is the hopping
matrix between two successive layers. The overlap matrix $\bm{S}_{\alpha}$ has the
same form. $\bm{H}_{0}$ and $\bm{H}_{1}$ are matrices of size $N_{\alpha}$ where $N_{\alpha}$ is the
number of atomic spin-orbitals in a layer of the lead $\alpha$.

We need to calculate the surface Green function only on the atoms that have interactions
with the scattering region. Consequently we just have to calculate the surface
Green function on the last layer. Thus $\bm{g}_{\alpha}^{S}$ is a matrix of size $N_{\alpha}\times N_{\alpha}$
defined by the relation:

\begin{equation}
\begin{split}
&
\left(\begin{array}{cccc}
\ddots & \ddots        & \ddots       & \vdots        \\
\ddots & E\bm{S}_{0} - \bm{H}_{0}         & E\bm{S}_{1} - \bm{H}_{1}        & 0             \\
\ddots & E\bm{S}_{1}^{\dag} - \bm{H}_{1}^{\dag}  & E\bm{S}_{0} -  \bm{H}_{0}        & E\bm{S}_{1} - \bm{H}_{1}
        \\
\ldots & 0             & E\bm{S}_{1}^{\dag} - \bm{H}_{1}^{\dag} &  E\bm{S}_{0} - \bm{H}_{0}        
 \\  
\end{array}\right) \times
\\
&
\left(\begin{array}{cccc}
\ddots  & \ddots   & \vdots	       & \vdots           \\
\ddots  & \ldots   & \ldots            & \ldots              \\
\ldots  & \ldots   & \bm{G}_{\alpha}^{S-1}  & \bm{G}_{\alpha}^{S-1S}   \\  
\ldots  & \ldots   & \bm{G}_{\alpha}^{SS-1} & \bm{g}_{\alpha}^{S}    \\  
\end{array}\right) 
=
\left(\begin{array}{cccc}
\ddots  & \ddots   & \vdots  & \vdots  \\
\ddots  & \bm{I}_{N_{\alpha}}        & 0       & 0  \\
\ldots  & 0        & \bm{I}_{N_{\alpha}}      & 0   \\  
\ldots  & 0        & 0       & \bm{I}_{N_{\alpha}}   \\  
\end{array}\right) .
\end{split}
\label{GL}
\end{equation}

Now we add a layer (denoted as $S+1$) to the system. Using the Dyson equation, we
can calculate the surface Green function $\bm{g}_{\alpha}^{S+1}(E)$ of the new system
from the surface Green function $\bm{g}_{\alpha}^{S}$ of the old system and the
Hamiltonians $\bm{H}_{0}$ and $\bm{H}_{1}$:   

\begin{equation}
\begin{split}
& \bm{g}_{\alpha}^{S+1}(E)= \\
& \big(E\bm{S}_{0} - \bm{H}_{0} - (E\bm{S}_{1}^{\dag} -
\bm{H}_{1}^{\dag})\bm{g}_{\alpha}^{S}(E)(E\bm{S}_{1} - \bm{H}_{1})\big)^{-1}
\end{split}
 \label{sgf+1}
\end{equation}

Due to the periodicity of the system, adding a layer does not change the
Green function on the surface layer. 
Therefore, $\bm{g}_{\alpha}^{S+1}(E)=\bm{g}_{\alpha}^{S}(E)$ and the surface Green
function is solution of the equation

\begin{equation}
\begin{split}
& \bm{g}_{\alpha}^{S}(E)=\\
& \big(E\bm{S}_{0} - \bm{H}_{0} - (E\bm{S}_{1}^{\dag} 
- \bm{H}_{1}^{\dag})\bm{g}_{\alpha}^{S}(E)(E\bm{S}_{1} - \bm{H}_{1})\big)^{-1}
\end{split}
 \label{sgf}
\end{equation}

This equation is solved iteratively using the quick iterative scheme proposed by
L\'opez Sancho {\sl et al.}\cite{Lopez84}.

\section{Transmission of a finite monoatomic wire connected to two semi-infinite wires \label{sec:finite_wire}}

We consider two identical semi-infinite linear chains (leads) connected to a finite chain of $N$
atoms of a different species (scattering region) (See Fig. \ref{fig:geom}a).  All the atoms 
are non magnetic and equally
spaced with an interatomic distance taken as unity. An $s$ atomic orbital is centered
on each atom. The electronic states are obtained from a tight-binding scheme in
which the non-orthogonality of the $s$ orbitals is neglected. Furthermore all the atomic
levels of the system are assumed to be the same and taken as the energy zero. The hopping integrals
are limited to first nearest neighbours and are equal to $\beta_0$ and $\beta$ in
the leads and scattering region, respectively. Finally the connection between the
leads and the scattering region is established by an hopping integral $\gamma$
(See Fig. \ref{fig:geom}a). The Schroedinger equation projected on each
atomic site yields (for each spin):

\begin{eqnarray}
 \beta_0a_{n-1}-Ea_n+\beta_0a_{n+1}      = 0   & &\quad n\le -1 \label{eq1}    \\
 \beta_0a_{-1}-Ea_0+\gamma a_{1}         = 0   & &   \label{eq2} \\              
 \gamma a_{0}-Ea_1+\beta a_{2}          = 0   & &   \label{eq3}  \\           
 \beta a_{n-1}-Ea_n+\beta a_{n+1}        = 0   & &  \quad 2<n<N-1 \;  \  \; \  \label{eq4}\\    
 \beta a_{N-1}-Ea_N+\gamma a_{N+1}       = 0   & &         \label{eq5} \\           
 \gamma a_{N}-Ea_{N+1}+\beta_0 a_{N+2}  = 0   & &       \label{eq6} \\            
 \beta_0a_{n-1}-Ea_n+\beta_0a_{n+1}      = 0   & & \quad n>N+1  \label{eq7}   
\end{eqnarray}

\noindent
$a_n$ being the coefficient of the wave function relative to the $s$ atomic orbital centered on atom $n$.
We look for a solution in which an electronic wave with crystal momentum $k_0>0$ inside the left lead is reflected by
the defect with a reflection amplitude probability $r$  and transmitted into the
right lead with a transmission amplitude probability $t$.
Then the amplitude of the electronic wavefunction on each atom can be written:

\begin{equation} \label{eq:cases}
a_n=\begin{cases} e^{ink_0}+re^{-ink_0} & \text{if $n \leq 0$ },\\ 
                  Ae^{ink}+Be^{-ink}    & \text{if $1 \le n \leq N$ },\\
                  te^{ink_0}            & \text{if $n > N$}. \end{cases}
\end{equation}

\noindent
which obviously satisfies Eqs \ref{eq1},\ref{eq4} and \ref{eq7} with 
$E=2 \beta_0 \cos k_0 =2 \beta \cos k $ (Note that $k$ may be imaginary when
$|\beta|<|\beta_0|$). Substituting \ref{eq:cases} for $a_n$ into Eqs \ref{eq2},
\ref{eq3}, \ref{eq5} and \ref{eq6} yields a linear system of 4 equations with the
4 unknown variables $A$, $B$, $r$ and $t$. After solving this system the transmission factor
is $T_N(E)=2tt^*$ given as a function of $\beta$, $\beta_0$, $\gamma$ and $N$ by Eq. \ref{eq:T_N}
of the main text.

\end{document}